\documentclass{ws-procs975x65}

\def\beq{\begin{equation}}
\def\eeq{\end{equation}}

\begin{document}

\title{Soft gamma-ray repeaters and anomalous X-ray pulsars as highly magnetized white dwarfs}
\author{Banibrata Mukhopadhyay}

\address{Department of Physics, Indian Institute of Science, Bangalore 560012\\
E-mail: bm@physics.iisc.ernet.in\\
}

\author{A. R. Rao} 

\address{Tata Institute of Fundamental Research, Mumbai 400005, India\\
E-mail: arrao@tifr.res.in
}

\begin{abstract}
We show that the soft gamma-ray repeaters (SGRs) and anomalous X-ray pulsars (AXPs)
can be explained as recently proposed highly magnetized white dwarfs (B-WDs). The radius
and magnetic field of B-WDs are perfectly adequate to explain energies in SGRs/AXPs as
the rotationally powered energy.
While the highly magnetized neutron stars
require an extra, observationally not well established yet, source of energy,
the magnetized white dwarfs, yet following Chandrasekhar's theory (C-WDs),
exhibit large ultra-violet luminosity which is observationally constrained from a strict upper limit.
\end{abstract}

\keywords{white dwarfs; stars: magnetars; stars: magnetic field;
gravitation; relativity}

\bodymatter


\section{Introduction}\label{intro}

Soft gamma repeaters (SGRs) and anomalous X-ray pulsars (AXPs) are, 
as of now, most popularly hypothesized as isolated, spinning down, highly magnetized neutron stars (NSs)
(magnetar model) \cite{thomdun92}. 
A NS of radius $10$km with surface and central magnetic fields respectively $B_s\sim 10^{14}$G and $B_c\sim 10^{16}$G
can have magnetic energy $\sim 10^{48}$erg, which could produce the luminosity
$\sim 10^{36}$erg/sec, as observed for AXPs/SGRs, in its typical age. 
However, there are certain shortcomings in it. First of all, as of now, there
is no evidence for a strongly magnetized NS --- as strong as required for the magnetar
model.
Second, recent Fermi observations
are inconsistent with predicted high energy gamma-ray emissions in the magnetars.
Third, inferred upper limit of $B_s$ for some magnetars, e.g. SGR~0418+5729,
is quite smaller than the field required to explain observed X-ray luminosity.
Fourth, the attempt to relate magnetars to the energies of supernova remnants or
the formation of black holes is not viable. There are many more.
These observations imply that the high magnetic dipole moment is not a mandatory condition
for a magnetar.

Recently AXPs/SGRs have been argued \cite{mane} to be magnetized white dwarfs (WDs), following the
idea proposed decades back \cite{pac,ost}. Due to their larger radius, 
the rotationally powered energy for WDs
could be quite larger than that for NSs. Hence, these authors attempted
to explain the energy released in AXPs/SGRs through the occurrence of glitch and
subsequent loss of the rotational energy. 
While this WD based
model (hereinafter C-WD) does not need to invoke extraordinarily strong, unconfirmed observationally yet,
magnetic field, it is challenged by the observed short
spin periods (e.g. Ref.~\refcite{meregh13}). In addition, due to larger radius, they should exhibit larger
ultra-violet (UV) luminosities,
which however, suffer from a deep upper limits on the optical counterparts (e.g. Refs.~\refcite{meregh13,durant}) of
some AXPs/SGRs, e.g. SGR~0418+5729.

Recently, Mukhopadhyay and his collaborators, in a series of papers, have proposed
for the existence of highly magnetized WDs (see, e.g., Refs.~\refcite{prd,prl,apjl,jcap1,jcap2})
with mass significantly super-Chandrasekhar. 
$B_s$ of such WDs
could be as high as $10^{12}$G and $B_c$ could be $2-3$ orders of magnitude higher.
These WDs (hereinafter B-WDs) are significantly
smaller in size compared to their ordinary counterparts (e.g. polar with $B_s\sim 10^9$G).
Their radius could even be an order or order and half of magnitude higher than that of a NS.
As the surface temperatures of WDs with different magnetic fields
are not expected to differ significantly \cite{magobs}, smaller the radius, smaller the luminosity
of the WD is. Therefore, B-WDs should be consistent
with the UV-luminosity ($L_{UV}$) cut-off in AXPs/SGRs. Moreover, their typical $B_s$ is consistent with observations,
but adequate to explain AXP/SGR energies as rotationally/spin-down powered energy, unlike the NS based
models and C-WDs.

Here we explore AXPs/SGRs as B-WDs. Although the evolution of B-WDs was argued by accretion, they may
appear as AXPs/SGRs at the exhaustion of mass supply after significant evolution.
Such WDs' $B_s$ and $R$ combination can easily explain AXPs/SGRs as
rotationally powered WDs.
All the machineries implemented in the magnetar model
can be applicable for B-WDs as well, however, with a smaller $B_s$ which
is physically more viable. Hence, under the B-WD model, one does not necessarily need to
invoke an extra-ordinarily source of magnetic energy --- everything comes out naturally.
We also show that the magnetic fields in B-WDs are in accordance with the virial theorem,
subject to its modification based on the magnetic pressure in the magnetostatic condition.

\section{Modelling magnetized white dwarfs as rotating dipoles}\label{data}

The rate of energy loss
from an oscillating magnetic dipole is 
\begin{equation}
\dot{E}_{\rm rot}=-\frac{\mu_0\Omega^4\sin^4\alpha}{5\pi c^3}|m|^2,
\label{edot}
\end{equation}
when the variation of dipole moment $m$ arises due to a magnetic dipole having inclination
angle $\alpha$ with respect to its rotational axis, $\Omega$ is the angular frequency
of the dipole, $c$ the light speed. Assuming the rotating, magnetized compact stars 
to be rotating magnetic dipole, the dipole nature of magnetic field
is expressed as
\begin{equation}
B=\frac{\mu_0 |m|}{2\pi R^3},
\end{equation}
when $R$ is the radius of star. However, the above energy loss rate can be defined as
the rate of rotational kinetic energy change of star with moment of inertia $I$ as
\begin{equation}
I\dot{\Omega}=\dot{E}_{\rm rot},
\end{equation}
which leads to
\begin{equation}
B_s=\sqrt{\frac{15c^3IP\dot{P}}{\pi^2R^6\sin^2\alpha}}~G,
\label{bs}
\end{equation}
when $P$ is the rotational period and $\dot{P}$ the period derivative, $I=Mk^2$,
$M$ the mass of the compact star. Note that $k\propto R$ and the proportionality
constant depends on the nature of the matter and its distribution in the star and shape of the star.
This is the upper limit of $B_s$.
As $P$ and $\dot{P}$ for AXPs/SGRs are known from observations, $B_s$ can be computed for
a given mass-radius ($M-R$) relation when $\alpha$ is a parameter. Once $B_s$ is estimated for
an observation, the rotational/dipole energy $E_{\rm rot}$ stored in the star can be computed. This
further quantifies the maximum energy stored in it, if there is no other source as adopted
in the magnetar model.



\section{Explaining AXPs/SGRs}\label{result}

We consider nine AXPs/SGRs
explained as B-WDs,
listed in Table~1. 
We also assume the surface temperature
of WDs to be $T_{UV}\sim 10^4$ K 
with WDs to be
semi-solid sphere/ellipsoid.



\begin{figure}[h]
\begin{center}
\includegraphics[angle=270,width=4in]{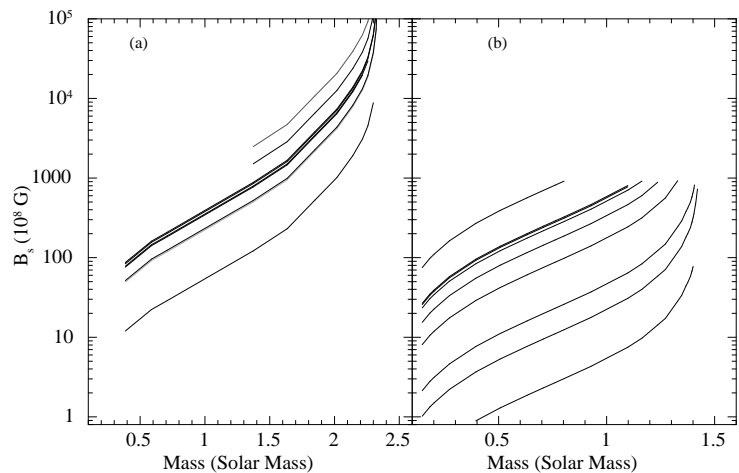}
\end{center}
\caption{
Surface magnetic field as a function of mass for (a) B-WDs when
from the top to bottom various curves correspond to SGR~1806-20, 1E~1048-59,
SGR~0526-66, 1E~1841-045, 1E~1547-54, SGR~1900+14, 1E~2259+586, SGR~1822-1606, SGR~0418+5729, (b) C-WDs
when from the top to bottom various curves correspond to SGR~1806-20, SGR~0526-66,
1E~1841-045, SGR~1900+14, 1E~1048-59, 1E~1547-54, 1E~2259+586, SGR~1822-1606, SGR~0418+5729.
}
\label{fig}
\end{figure}

\begin{table}
\tbl{Various observational and theoretical parameters of AXPs/SGRs: $P$, $\dot{P}$,  $L_x$ are
observed values and inputs and $\alpha$, minimum of $L_{UV}$ are outputs of our model.
See, http://www.physics.mcgill.ca/$\sim$pulsar/magnetar/main.html}
{\begin{tabular}{@{}ccccccc@{}}
\toprule
\hline
AXPs/SGRs & $P$ &  $\dot{P}$ & $L_x$ & $\alpha$
& $L_{UV}$$_{\rm min}$ & $L_{UV}$$_{\rm min}$ \\
 & (sec) & $(10^{-11})$ & $(10^{35}{\rm erg/sec})$ & (degree) & (erg/sec)& erg/sec\\
 & &  &  &  &B-WD & C-WD\\
\hline
1E~1547-54 & $2.07$ & $2.32$ & $0.031$ & $5-15$ & $5.7\times10^{28}$&$4.8\times10^{29}$\\
\hline
1E~1048-59 & $6.45$ & $2.7$ & $0.054$ & $5-15$ & $3.5\times10^{26}$& $9.2\times10^{29}$\\
\hline
1E~1841-045 & $11.78$ & $4.15$ & $2.2$ & $15$ &$1.6\times10^{28}$&$1.7\times10^{30}$\\
\hline
1E~2259+586 & $6.98$ & $0.048$ & $0.19$ & $2-3$& $3.4\times10^{26}$ &$1.5\times10^{29}$\\
\hline
SGR~1806-20 & $7.56$ & $54.9$ & $1.5$ & $15$ &$3.4\times10^{26}$&$3.5\times10^{30}$\\
\hline
SGR~1900+14 & $5.17$ & $7.78$ & $1.8$ & $15$ &$8.6\times10^{28}$ & $1.3\times10^{30}$\\
\hline
SGR~0526-66 & $8.05$ & $6.5$ & $2.1$ & $15$ &$6.4\times10^{27}$ & $1.7\times10^{30}$\\
\hline
SGR~0418+5729 & $9.08$ & $5\times 10^{-4}$ & $6.2\times 10^{-4}$ & $1-5$ &$3\times10^{28}$ &$1.8\times10^{29}$\\
\hline
SGR~1822-1606 & $8.44$ & $9.1\times 10^{-3}$ & $4\times 10^{32}$ & $1-5$ &$3.4\times10^{26}$ &$8\times10^{28}$\\
\hline
\botrule
\end{tabular}}
\label{aba:tbl1}
\end{table}

Figures \ref{fig}a and \ref{fig}b show that $B_s$ for the $M-R$ combination
of the B-WDs is quite stronger compared to that of the C-WDs. Also B-WDs exhibit X-ray luminosity:
$100\lesssim\dot{E}_{\rm rot}/L_x\lesssim 10^7$, 
from Eq. (\ref{edot}),
explaining all the AXPs/SGRs listed in Table~1 as rotational powered pulsars. 

This naturally explains AXPs/SGRs without requiring an extra-ordinary source of magnetic energy.
Generally, higher the $B_c and/or B_s$ for a poloidal dominated field, higher the $M$ is, which corresponds to a lower $R$
and hence a lower $L_{UV}$. 

\section{Maximum allowed magnetic field and modified virial theorem}\label{vir}

First note very importantly that in the presence of strong magnetic field, 
the upper limit of magnetic fields in WDs, as discussed in, e.g. Ref.~\refcite{shapiro} for weak field cases,
has to be revised, the contribution of the magnetic pressure to the 
hydro/magnetostatic balance equation cannot be neglected. Here we attempt to revise
so in a simpler/approximate framework.

If the gravitational, thermal and magnetic energies are respectively denoted by
$W$, $\Pi$ and $\mu$, then the scalar virial theorem can be read as
\begin{equation}
W+3\Pi+\mu=0,~~{\rm i.e.}~~-\alpha\frac{GM^2}{R}+3M\frac{P}{\rho}+\frac{B^2}{24\pi}~\frac{4}{3}\pi R^3, 
\label{v2}
\end{equation}
where we consider, on average, the isotropic effects of averaged magnetic field $B$ and hence overall the 
star to be spherical in shape, $P$ is the pressure of the stellar matter, $\rho$ the averaged density,
$M$ the mass of WD, $G$ the Newton's gravitation constant.
Now we assume that a polytropic equation of state (EoS) to be satisfied in entire star such that
$P=K\rho^\Gamma$, where $K$ and $\gamma$ are the polytropic constants, and
$M=\frac{4}{3}\pi R^3\rho$. Therefore, the scalar
virial theorem can be reduced to
\begin{equation}
-\alpha\frac{GM^2}{R}+\beta \frac{M^\Gamma}{R^{3(\Gamma-1)}}+\gamma\frac{\Phi_M^2}{R},
\label{vir}
\end{equation}
where $\Phi_M=B\pi R^2$ and $\alpha$, $\beta$ and $\gamma$ are the constant factors 
determined by the shape and other properties of the star.

Now combining first and second terms of Eq. (\ref{vir}), we obtain
\begin{eqnarray}
M=\sqrt{\frac{\gamma\phi_M^2}{\alpha G\left(1-\frac{\beta M^{\Gamma-2}}{\alpha G R^{3\Gamma-4}}\right)}},
\label{mas3}
\end{eqnarray}
which is valid for any value $\Gamma$. For $\Gamma=4/3$, it gives
$M=\sqrt{\frac{\gamma\phi_M^2}{\alpha G\left(1-\frac{\beta M^{-2/3}}{\alpha G}\right)}}$ 
which is independent of $R$, as expected from Chandrasekhar's theory, and 
needs to be solved for $M$. However, for $\Gamma=2$, this gives 
$M=\sqrt{\frac{\gamma\phi_M^2}{\alpha G\left(1-\frac{\beta}{\alpha G R^2}\right)}}$.
Conversely, one can write $R=\sqrt{\beta/\alpha G^\prime}$, where 
$G^\prime=G\left(1-\gamma\phi_M^2/\alpha GM^2\right)$.

Now in order to obtain the value of $M$ and $R$ explicitly, we have to evaluate the values
of $\alpha$, $\beta$ and $\gamma$. 
The magnetostatic balance condition 
is given by
\begin{equation}
\frac{1}{\rho}\frac{dP}{dr}+\frac{1}{\rho}\frac{dP_B}{dr}=-\frac{Gm(r)}{r^2}
\label{magst}
\end{equation}
at an arbitrary radius $r$ from the center of the star with mass enclosed at that radius $m(r)$,
where $\rho$ includes the contribution from $B$ as well.
We further assume the variation of $B$ to be a power law like such that the 
corresponding magnetic pressure 
$P_B=K_1\rho^\Gamma_1$ with $K_1$ and $\Gamma_1$ being constants. 
Hence, the gravitation energy for this star is
\begin{eqnarray}
\nonumber
W&=&-\int_0^R\frac{Gm(r)}{r}4\pi r^2dr\rho=\int_0^R 4\pi r^2dr\rho~\frac{r}{\rho}
\left(\frac{dP}{dr}+\frac{dP_B}{dr}\right)\\
&=&
-\frac{3(\Gamma_1-1)}{5\Gamma_1-6}\frac{GM^2}{R}+\frac{\Gamma-\Gamma_1}{5\Gamma_1-6}~\frac{3\Pi}{\Gamma-1},
\label{grave}
\end{eqnarray}
assuming that $\rho$ is negligibly small at $r=R$, surface of the star, compared to that at the center
(or its averaged value).
Therefore, from Eqs. (\ref{v2}) and (\ref{vir}), we obtain
\begin{eqnarray}
-\frac{3(\Gamma_1-1)}{5\Gamma_1-6}\frac{GM^2}{R}+\left(1+\frac{\Gamma-\Gamma_1}{(5\Gamma_1-6)(\Gamma-1)}
\right)3\Pi+\mu=0
\end{eqnarray}
and consequently
\begin{equation}
\alpha=\frac{3(\Gamma_1-1)}{5\Gamma_1-6},~~~\beta=\left(1+\frac{\Gamma-\Gamma_1}{(5\Gamma_1-6)(\Gamma-1)}
\right)\frac{K~3^\Gamma}{(4\pi)^{\Gamma-1}},~~~\gamma=\frac{1}{18\pi^2}.
\end{equation}

An important outcome here is that $\alpha$
is related to the scaling of $B$ with $\rho$, which is indeed expected 
from the magnetostatic balance Eq. (\ref{magst}). In other words, the presence of magnetic 
pressure allows either a massive or/and smaller star. 
Obviously, for $\Gamma=\Gamma_1$ the 
result reduces to that of the nonmagnetic case with the redefinition $K$.

Following Refs.~\refcite{shapiro,mane}, the upper bound of $B$ for a gravitationally bound star corresponds to
\begin{eqnarray}
\mu=\frac{B^2 R^3}{18}=\frac{3(\Gamma_1-1)}{5\Gamma_1-6}\frac{GM^2}{R}
\end{eqnarray}
which leads to maximum allowed $B$
\begin{eqnarray}
B_{max}=\sqrt{\frac{54(\Gamma_1-1)}{(5\Gamma_1-6)}\frac{GM^2}{R^4}}=7.829\times 10^8\frac{M}{M_\odot}
\left(\frac{R_\odot}{R}\right)^2 \sqrt{\frac{\Gamma_1-1}{5\Gamma_1-6}} {\rm G},
\label{bmax}
\end{eqnarray}
where $R_\odot$ is the radius of Sun. For a B-WD having $B_c=8.8\times 10^{15}$G, 
and hence averaged $B_{max}=4.4\times 10^{15}$G, $M=2.44$ solar mass and $R=654$km, 
with $\Gamma=2$, as reported in Refs.~
\refcite{prd,prl}, in order to satisfy Eq. (\ref{mas3}), $\Gamma_1$ has to be $1.2029$.
Similarly, for the case of $M=1.77$ solar mass, equatorial $R=956.14$km, $B_c=5.34\times 10^{14}$G,
with $\Gamma\approx 4/3$ (Ref. \refcite{jcap2}), $\Gamma_1$ has to be $3.5589$.
Importantly, $P(\rho)$ profile and, hence, $\Gamma$ is determined by $P_B(\rho)$ profile, which 
however has not been strictly followed in this approximate calculation.

\section{Conclusions}\label{conc}

The present work indicates a possibility of
wide application of recently proposed B-WDs in modern astrophysics. The idea that AXPs/SGRs need sources
of energy other than rotational or accretion is certainly inevitable, but the hypothesis
that they are highly magnetic NSs, although attractive, did not neatly fit in
with further observations (unlike other ideas in Astrophysics, like, spinning NSs
as radio pulsars and accreting compact objects as X-ray binaries, which quickly
established themselves as paradigms). Hence, it
is very important that other possible explanations for the AXP/SGR phenomena need to be
seriously explored. The B-WD concept is an extremely attractive alternate for AXPs/SGRs.

\section*{Acknowledgments}
We thank Upasana Das, our coauthor of many papers in this topic, for discussion.

\end{document}